\documentclass[12pt]{article}
\usepackage{epsf}
\def\be{\begin{equation}}
\def\ee{\end{equation}}
\def\bea{\begin{eqnarray}}
\def\eea{\end{eqnarray}}

\begin{document}
\begin{titlepage}
\begin{center}
{\Large \bf William I. Fine Theoretical Physics Institute \\
University of Minnesota \\}
\end{center}
\vspace{0.2in}
\begin{flushright}
FTPI-MINN-14/16 \\
UMN-TH-3340/14 \\
July 2014 \\
\end{flushright}
\vspace{0.3in}
\begin{center}
{\Large The decays $\Xi_b \to \Lambda_b \, \pi$ and diquark correlations in hyperons.
\\}
\vspace{0.2in}
{\bf Xin Li$^a$  and M.B. Voloshin$^{a,b,c}$  \\ }
$^a$School of Physics and Astronomy, University of Minnesota, Minneapolis, MN 55455, USA \\
$^b$William I. Fine Theoretical Physics Institute, University of
Minnesota,\\ Minneapolis, MN 55455, USA \\
$^c$Institute of Theoretical and Experimental Physics, Moscow, 117218, Russia
\\[0.2in]

\end{center}

\vspace{0.2in}

\begin{abstract}
The decays $\Xi_b \to \Lambda_b \pi$ are strangeness changing weak transitions involving only the light diquark in the baryon. Thus these decays can test the properties of such diquarks, in particular the suggestions existing in the literature of enhanced correlations in $J^P=0^+$ light diquarks. We revisit the estimates of the rates of these decays and point out that with the enhanced correlation their branching fraction can reach a few percent and may become visible in the measurements of differences of the lifetimes of $b$ baryons.   
\end{abstract}
\end{titlepage}

The lifetime of weakly decaying heavy hadrons is dominantly determined by the decay of the heavy quark. However the strange heavy hyperons can also decay without the destruction of the heavy quark, but rather due to the weak decay of the strangeness~\cite{sk,mv00}, e.g. $\Xi_b^0 \to \Lambda_b \, \pi^0$ and $\Xi_b^- \to \Lambda_b \, \pi^-$. The rate of these decays is expected to be in the same ballpark as that of the ordinary strange hyperons, i.e. of the order of 0.01\,ps$^{-1}$. Furthermore, the heavy baryons are likely to be more spatially compact systems than the light ones, so that a certain enhancement of the weak pion transitions for the heavy hyperons would be quite natural. It is expected that such decays contribute only at a permille level to the lifetimes of the charmed hyperons, while their contribution to the total decay rate of the strange $b$ baryons can be essential at a level of one or few percent and their effect on the lifetimes of $\Xi_b^0$ and $\Xi_b^-$ may become visible once the experimental data become sufficiently precise. Thus the strangeness changing weak transitions could potentially affect comparison of the data with the results based on the Heavy Quark Expansion (HQE) for the dominant processes generated by the $b$ quark decay. The purpose of the present paper is to point out that in addition to a possible relevance to studies of the HQE, the weak strangeness changing pion transitions for the $b$ hyperons are interesting on their own and provide a testing ground for  the properties of $J^P=0^+$ light diquarks which, according to some theoretical suggestions (e.g. in Ref.~\cite{sv}), can contain a large QCD scale that would enhance the rate of such transitions. 

The lifetimes of the weakly decaying $b$ baryons, $\Lambda_b$, $\Xi_b^0$ and $\Xi_b^-$, are traditionally a subject of a significant  interest in connection with the HQE calculations for the inclusive decay rates of heavy hadrons. 
Recent improvements in the accuracy and reliability of the experimental data on these lifetimes~\cite{stone} have to a great extent resolved the tensions of the early LEP measurements of $\tau (\Lambda_b)$ with the predictions from HQE. Namely, the latest data with the highest reported precision by the LHCb experiment~\cite{lhcb1,lhcb2} result in the average value~\cite{stone}  $\tau (\Lambda_b) = 1.468 \pm 0.009 \pm 0.008$\,ps, and the recent results, albeit of a lower accuracy, from other experiments~\cite{cdf,d0,cms,atlas} are consistent with this value. These measurements place the ratio of the lifetimes $\tau(\Lambda_b)/\tau(B_d^0)$ well inside the range 0.95 - 1.0, in agreement with the original HQE prediction~\cite{sv86} as well as with the more recent theoretical estimates~\cite{lenz}. Now, as the $\Lambda_b$ lifetime issue between the HQE and the experiment is apparently resolved, the theory can be put to further test against the currently improving data~\cite{cdf,lhcb3,lhcb4} on the lifetimes of the rest of the weakly decaying $b$ baryons, which may also provide a quantitative insight in the internal structure of baryons. In this respect of a special interest are the differences of the inclusive decay rates within the flavor SU(3) (anti)triplet of the $b$ baryons $\Lambda_b$, $\Xi_b^0$ and $\Xi_b^-$ as well as their comparison with similar differences in the antitriplet of the charmed hyperons $\Lambda_c$, $\Xi_c^0$ and $\Xi_c^+$.

The HQE predicts a very small difference between the inclusive rates of the beauty decay in the $\Lambda_b$ and $\Xi_b^0$ hyperons, similar to the prediction of a less than one percent difference between the lifetimes of the $B_d^0$ and $B_s^0$ mesons, due to the approximate (but in fact quite accurate) $U$ spin symmetry of the leading term in HQE that depends on the flavors of the spectator light quarks~\footnote{A discussion of this property can be found e.g. in Ref.~\cite{lenz}.}. The current data~\cite{lhcb3} do not contradict this prediction: $\tau(\Xi_b^0)/\tau(\Lambda_b) = 1.006 \pm 0.018 \pm 0.010$. However it is also not excluded, within the errors, that the $\Xi_b^0$ decay rate receives an extra few percent contribution from the weak transition $\Xi_b^0 \to \Lambda_b \, \pi^0$.

It is also expected~\cite{sv86} from HQE that the lifetime of $\Xi_b^-$ is noticeably longer than that of $\Lambda_b$. A quantitative prediction depends on hadronic matrix elements of the four-quark operators arising in the expansion for the inclusive weak decay rate which matrix elements were studied purely theoretically~\cite{codf,ukqcd} and also phenomenologically~\cite{mv99} from the measured lifetime differences for the charmed baryons. The latter determination of the relevant baryonic matrix elements results in the estimated difference between the inclusive beauty decay rate for $\Lambda_b$ and $\Xi_b^-$ equal to $0.11 \pm 0.03$\,ps$^{-1}$. This value is substantially larger than the two available experimental numbers for the difference of total decay rates for these $b$ baryons: the reported CDF result~\cite{cdf}, $\tau(\Xi_b^-)=1.32 \pm 0.14 \pm 0.02 \,$ps gives the opposite sign for the difference compared to HQE prediction, while the latest LHCb number $\tau(\Xi_b^-)=1.55^{+0.10}_{0.09} \pm 0.03 \,$ps corresponds to $\tau^{-1} (\Lambda_b) - \tau^{-1}(\Xi_b^-)=0.03 \pm 0.04 \,$ps$^{-1}$, and is formally compatible with the expected difference of the inclusive decay rates only due to the large experimental errors and the theoretical uncertainty. 

Given that the experimental situation with the lifetimes of the strange $b$ baryons is still in flux, one can consider several options. One is that an application of the leading terms in HQE to the charmed hyperons is not quantitatively reliable due to insufficiently heavy mass of the charmed quark, so that the estimate~\cite{mv99} of the relevant baryonic matrix elements from the lifetime differences of the charmed hyperons results in an exaggerated prediction for the difference of the beauty decay rates between $\Lambda_b$ and $\Xi_b^-$. Although this is not excluded, there are arguments (see e.g. in Ref.~\cite{lenz}) against such reasoning. Another possibility, based on the history of measurements of $\tau(\Lambda_b)$, is that future more accurate experimental data will eventually settle at a longer lifetime for the $\Xi_b^-$ in agreement with the HQE estimate in Ref.~\cite{mv99}. It is however also possible that the actual value of $\tau(\Xi_b^-)$ is somewhere half way between the current data and the theoretical expectation, and the strangeness changing weak pion transitions are enhanced in comparison with the ordinary hyperons, so that there is a noticeable addition (at the level of few percent) to the total decay rate of the $\Xi_b$ hyperons. It can be noticed that in the decays $\Xi_b \to \Lambda_b \, \pi$ the $\Delta I = 1/2$ rule should hold to a high accuracy~\cite{mv00}, so that
\be
\Gamma(\Xi_b^- \to \Lambda_b \, \pi^-) = 2 \, \Gamma(\Xi_b^0 \to \Lambda_b \, \pi^0)~,
\label{di12}
\ee
and an additional decay rate for $\Xi_b^0$ being still compatible with the current data may translate to a few percent shortening of the lifetime of $\Xi_b^-$. Certainly, a direct observation of the transitions $\Xi_b \to \Lambda_b \, \pi$ would definitely solve the issue, however it is not clear whether such observation is feasible in the LHC environment.

In the strangeness changing weak transitions between $b$ baryons the heavy $b$ quark is a spectator, and the process is determined entirely by dynamics of the light diquark which has quantum numbers $J^P=0^+$. It can be mentioned, that a mixing with the $b$ baryons from the flavor SU(3) sextet in which the light diquark is in the $J^P=1^+$ state is suppressed by both the flavor SU(3) symmetry and by the heavy quark spin symmetry, so that the suppression factor in the mixing amplitude is $O(m_s/m_b)$, and considering the diquark in a pure $0^+$ state makes a very good approximation.  In this limit the discussed decays of strangeness  are thus $0^+ \to 0^+$ transitions with emission of a pion, for which only the parity violating $S$ wave amplitude is allowed~\cite{mv00}. This amplitude can be approximated by its value at vanishing pion momentum, where it is given by the current algebra reduction formula 
\be
\langle \Lambda_b \, \pi_i (p=0) \,| H_W |\, \Xi_b \rangle = {\sqrt{2}
\over f_\pi} \, \langle \Lambda_b \, |\left[Q^5_i, \, H_W \right ] |\,
\Xi_b \rangle~,
\label{pcac}
\ee
where $H_W$ is the weak interaction Hamiltonian, $\pi_i$ is the pion triplet in the Cartesian notation, and $Q^5_i$
is the corresponding isotopic triplet of axial charges. The constant
$f_\pi \approx 130 \, MeV$, normalized by the charged pion decay, is
used here, which results in the the coefficient $\sqrt{2}$ in Eq.(\ref{pcac}). Clearly, the expression in the r.h.s. in Eq.(\ref{pcac}) is nonvanishing only for the part of the Hamiltonian corresponding to $\Delta I = 1/2$ which gives the relation (\ref{di12}). 

The strangeness changing weak Hamiltonian is well studied. At a normalization point $\mu$ below the charmed quark mass, $\mu \ll m_c$ this Hamiltonian has the general form~\cite{vzs}
\be
H_W=\sqrt{2} \, G_F \, \cos \theta_c \, \sin \theta_c \, \sum_{i=1}^{7} \, C_i (\mu) \, O_i(\mu)~,
\label{opeh}
\ee
where $\theta_c$ is the Cabibbo angle, and the explicit form of the operators $O_i$ can be found e.g. in Ref.~\cite{vzs}. In the present discussion the main part is being played by the operator  
\be
O_1=(\overline u_L \, \gamma_\mu \, s_L)\, (\overline
d_L \, \gamma_\mu \, u_L) - (\overline d_L \,
\gamma_\mu \, s_L)\, (\overline u_L \, \gamma_\mu \, u_L)~.
\label{o1}
\ee
Indeed, this is the only $\Delta I=1/2$ operator having a large coefficient in the expansion (\ref{opeh}): at a low $\mu$ one finds~\cite{vzs} $C_1 \approx 2.5$. Other $\Delta I = 1/2$ operators in the sum in Eq.(\ref{opeh})  have small coefficients (less than 0.1) and their matrix elements are not especially enhanced. For this reason, at the present level of accuracy, it is sufficient to retain in the expansion for the Hamiltonian only the term with the operator $O_1$. 

Using the reduction formula (\ref{pcac}), one readily finds the expression for the amplitude of the transition $\Xi_b^- \to \Lambda_b \, \pi^-$ in the form
\be
\langle \Lambda_b \, \pi^- (p=0) \,| H_W |\, \Xi_b^- \rangle =
{\sqrt{2} \over f_\pi} \, G_F \, \cos \theta_c \, \sin \theta_c \, C_1 \, X~,
\label{ampx}
\ee
with $X$ being the hadronic matrix element
\be
X=\langle \Lambda_b
\,| (\overline u_L \, \gamma_\mu
\, s_L)\, (\overline d_L \, \gamma_\mu \, d_L) -  (\overline d_L \, \gamma_\mu \,
s_L)\, (\overline u_L \, \gamma_\mu \, d_L)  | \,
\Xi_b^- \rangle~.
\label{xib}
\ee
Using the nonrelativistic normalization condition for the states of heavy baryons, one can write the rate of the pion transition in terms of the amplitude $X$ as
\be
\Gamma(\Xi_b^- \to \Lambda_b \, \pi^-) = \cos^2 \theta \sin^2 \theta \, C_1^2 \, {G_F^2 \, |X|^2 \, p_\pi \over \pi \, f_\pi^2} \approx 1.3 \times 10^{-2} \, {\rm ps}^{-1} \, \left ( {C_1 \over 2.5} \right )^2 \, \left | {X \over 0.01\, {\rm GeV^3} } \right |^2~,
\label{gx}
\ee 
where $p_\pi \approx 100$\,MeV is the pion momentum. Correspondingly, the expression for the branching fraction in terms of $X$ reads as
\be
{\cal B} (\Xi_b^- \to \Lambda_b \, \pi^-) \approx 2 \% \, \left ( {C_1 \over 2.5} \right )^2 \, \left | {X \over 0.01\, {\rm GeV^3} } \right |^2~.
\label{bx}
\ee

It has been suggested in the literature~\cite{djs,sv} that the scalar diquarks may have special properties resulting in an enhancement of short-distance correlations inside them similar to the matrix element $X$. One theoretical argument~\cite{sv} for such behavior is that in  QCD with two colors instead of three, and in the chiral limit the (colorless) diquark would be a Goldstone boson and exist on equal footing with a pion. Thus their respective decay constants should be the same. Extending this behavior to the actual QCD with three colors one can introduce\cite{djs} the `diquark decay constant' $g_D$ for a color-antitriplet diquark $D_i$ (e.g. made of $u$ and $d$ quarks),
\be
\langle 0 \, | \epsilon_{ijk} (\overline {u^c}^j \gamma_5 d^k) | \, D_l \rangle = \sqrt{2 \over 3} \, g_D \, \delta_{il}
\label{gd}
\ee
with $i,j,k,l$ being the triplet color indices, and compare it with an equivalent pion coupling to the pseudoscalar density
\be
\langle 0 \, | (\overline u \gamma_5 d) |\, \pi \rangle = g_\pi,~~~g_\pi = {f_\pi \, m_\pi^2 \over m_u+m_d} \approx 0.2\,{\rm GeV}^2~.
\label{gp}
\ee
(The factor $\sqrt{2/3}$ in Eq.(\ref{gd}) accounts for the different number of colors contributing in a pion and in a diquark of a fixed color.) If such extension from two colors to the actual three-color QCD is of relevance, one expects~\cite{djs} the approximate relation $g_D \approx g_\pi$.

Clearly, such diquark picture has intrinsically embedded the color antisymmetry:
\be
\langle  \Lambda_b
\,| (\overline u_L \, \gamma_\mu
\, s_L)\, (\overline d_L \, \gamma_\mu \, d_L)| \,
\Xi_b^- \rangle = - \langle \Lambda_b \, | (\overline d_L \, \gamma_\mu \,
s_L)\, (\overline u_L \, \gamma_\mu \, d_L)  | \,
\Xi_b^- \rangle~.
\label{as}
\ee 
Using then the Fierz identity $(\overline u_L \, \gamma_\mu
\, s_L)\, (\overline d_L \, \gamma_\mu \, d_L) = 2 \, (\overline d_{Li} u^c_{Rj}) \, ( \overline {s^c}^j_R d_L^i)$ one finds in the vacuum insertion dominance approximation
\be
X={g_D^2 \over 6 \, m_D}~,
\label{xg}
\ee
where $m_D$ stands for the `diquark mass'. Naturally, it is not quite clear what value should one use for $m_D$. However, the  study~\cite{djs} of the constant $g_D$ by the QCD sum rule method has produced a strong correlation between $m_D$ and $g_D$, such that the ratio $g_D^2/m_D$, entering Eq.(\ref{xg}) depends only weakly on the assumed value of $m_D$. Using Eq.(\ref{xg}) and the results presented in Ref.~\cite{djs}, we estimate 
\be
X \approx 0.01\,{\rm GeV}^3
\label{xgn}
\ee
This estimate understandably suffers from a considerable uncertainty, partly from the usage of speculative properties of $0^+$ diquarks, and partly, within this usage, from the usual uncertainties of the QCD calculation. In particular, the calculation of Ref.~\cite{djs} is carried out in the standard way and takes into account the first perturbative terms in the correlator of diquark densities and the terms with the quark and gluon vacuum condensates. However it has been pointed out later~\cite{sv} that such correlator should also receive a significant contribution from direct instantons, which would enhance the constant $g_D$, so that the actual value of the matrix element $X$ can be larger than estimated in Eq.(\ref{xgn}).

In order to probe how reasonable the value (\ref{xgn}) is, one can take an alternative approach and compare $X$ with similar four-quark matrix elements describing the light quark density at the heavy quark, rather than the density of light quarks at coinciding point:
\be
x=-\langle \Lambda_b \, | (\overline b \, \gamma_\mu \, b) (\overline u \,
\gamma_\mu \, s) | \, \Xi_b^- \rangle,~~~~~y= - \langle \Lambda_b \, | (\overline b_i \, \gamma_\mu \, b_k) (\overline
u_k \, \gamma_\mu \, s_i) | \, \Xi_b^- \rangle
\label{xyd}
\ee
The matrix elements $x$ and $y$ are related~\cite{mv00} by the heavy quark symmetry and the flavor SU(3) to the same quantities describing the lifetime differences of the charmed hyperons~\cite{mv99}. In a simplistic nonrelativistic picture these can be related~\cite{sv86} to the decay constant $f_B$ for the $B$ mesons: $y=-x = f_B^2 m_b/12 \approx 0.016$\,GeV$^3$. However, for these quantities the color antisymmetry relation $y=-x$ cannot be correct since it is not preserved by renormalization at $\mu \ll m_b$. Namely, the amplitude $y$  depends on the normalization scale $\mu$ in this range, while the amplitude $x$ does not. The amplitude $x$ can be determined~\cite{mv99} from the lifetime differences between $\Lambda_c,\, \Xi_c^0$, and $\Xi_c^+$, and the updated value is $x=-0.042 \pm 0.005$\,GeV$^3$, while the updated value of $y$ is small ($y < 0.01$\,GeV$^3$) across the range of $\mu$ below $m_c$. Unlike for the heavy-light correlators, the color antisymmetry for the light diquark densities in Eq.(\ref{as}) is preserved by renormalization. For this reason it is not quite clear how to compare the heavy-light and light-light correlations. In lieu of a better procedure, we replace the two heavy-light terms by their average, and find for the color antisymmetric combination $y-x \approx 0.03$\,GeV$^3$ both from the relation in the simplistic quark model and using the values extracted from the lifetimes of charmed hyperons. 

One may expect that the density of light quarks on the heavy quark as in $x$ and $y$ in a baryon is somewhat larger than the density of two light quarks on top of each other as in $X$. Indeed, the heavy quark is a static `center of force' at the `center' of the baryon, while the light quarks are each spread around inside the baryon. In order to estimate the numerical reduction factor for collisions between the light quarks as compared to the collisions of each with the center, we use the simple original bag model~\cite{cjjt}, and consider the heavy quark as static, i.e. in the limit of infinite mass. Using the wave functions for massless light quarks in the bag, we find
\be
X  \approx 0.62 \, (y-x) \approx 0.02\,{\rm GeV}^3~,
\label{bagx}
\ee
with the numerical factor arising from the ratio of the integrals over the quark wave functions: ${\int_0^{r_1} \left [ j_0^2(r)+ j_1^2(r) \right ]^2 \, r^2 \, {\rm d}r / \int_0^{r_1} \left [ j_0^2(r)+ j_1^2(r) \right ] \, r^2 {\rm d}\, r } \approx 0.62$, where $j_0(r)$ and $j_1(r)$ are the standard spherical Bessel functions, and $r_1 \approx 2.043$ is the smallest positive solution to $j_0(r_1) = j_1(r_1)$.

One can see, from the estimates presented here, that there is a considerable uncertainty in the value of the baryonic matrix element $X$ determining the rate of the weak decays $\Xi_b \to \Lambda_b \, \pi$. The range of these estimates $X \sim 0.01 \, - \, 0.02\,$GeV$^3$ corresponds to the branching fraction for such decay (2 - 8)\% for $\Xi_b^-$ and (1 - 4)\% for $\Xi_b^0$. At such level the contribution of these beauty-conserving decays can produce an effect on the lifetime differences between weakly decaying $b$ baryons that would be visible on top of the effects of HQE in the $b$ decay processes. Furthermore, since in these $b$ baryons the light quark is to a high accuracy a pure $0^+$ state, a measurement of the rates of these pion transitions and thus of the value of the matrix element $X$, may have very interesting implications for understanding the dynamics of $0^+$ diquarks. In particular, it would test the existing in the literature ideas about effects of short distance scales in such systems.

The work of M.B.V. is supported, in part, by the DOE grant DE-SC0011842.

\end{document}